\documentclass[12pt]{iopart}
\usepackage{bm}%
\usepackage{graphicx}
\usepackage{dcolumn}

\begin{document}
\title{Comment On \lq{}A modified Lennard-Jones-type equation of state for solids strictly satisfying the spinodal
condition\rq{}}
\author{A. Sai Venkata Ramana}
\address{Theoretical Physics Division, Bhabha Atomic Research Centre, Mumbai-400085, India}
\ead{asaivenk@barc.gov.in}

\begin{abstract}
The cohesive energies of solids calculated using mGLJ EOS proposed by Sun Jiuxun (Sun Jiuxun, J. Phys.: Condens. Matter 17,  L103 
(2005) ) are seen to be erroneous. Also we observed that the thermodynamic properties
 calculated using the mGLJ potential diverge for materials whose pressure derivative of bulk modulus at equilibrium is less
 than $5$. Thus the mGLJ potential cannot be used in liquid state theories and molecular simulations to obtain thermodynamic properties.
\end{abstract}
Sun Jiuxun \cite{Jiu} recently suggested an Equation of State(EOS) based on modified generalized Lennard-Jones (mGLJ) potential. 
The mGLJ EOS is obtained by modifying the generalized lennard-Jones potential in such a way that the EOS obtained is volume 
analytic and satisfies spinodal condition. The mGLJ EOS has three parameters and are related to lattice parameter, bulk modulus 
and derivative of the bulk modulus at equilibrium. 
 In the paper\cite{Jiu} it was shown that the pressure (P) Vs compression ratio curve obtained using the mGLJ EOS is quite accurate. 
The idea of generating an EOS starting from a potential is interesting and has the advantage that the potential of the material also can
 be known in addition to the EOS. 
However, we found some problems with the mGLJ EOS and the potential so obtained from it. 
They are, (a) the cohesive energy calculated from the mGLJ EOS is quite erroneous and (b) the thermodynamic properties calculated 
using the mGLJ potential with parameters of Sun diverge for most of the materials. Details about each problem are as follows:

\section {Cohesive Energy}
\begin{figure}
 {\includegraphics[scale=0.35,angle=270]{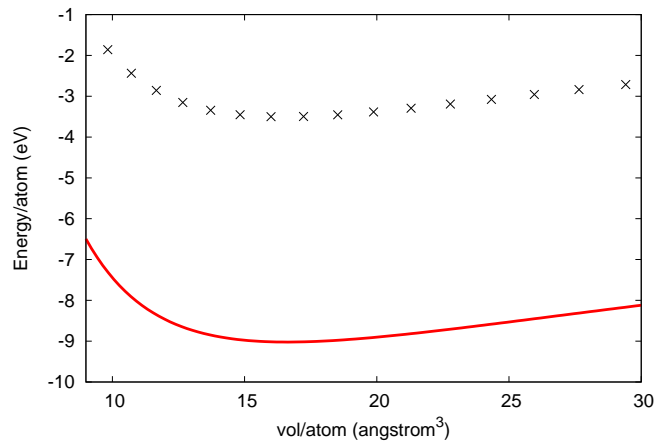}}%
 \caption{\label{1} Energy Vs Volume curve for Aluminum at temperature $T=0K$. Crosses are ab-initio data\cite{Vasp}. Solid line is obtained
 using Eq.(\ref{uv}) }
\end{figure}

The total energy formula obtained by Sun is
\begin{equation}
U(V) = \frac{9B_0V_0}{m_1n_1(m_1 - n_1)}\left[ n_1\left(\frac{V_0}{V}\right)^{m_1/3} - m_1\left(\frac{V_0}{V}\right)^{n_1/3} \right] \label{uv}
\end{equation} 
\noindent
where $U(V)$ is the total energy per atom, $B_0$ and $V_0$ are the equilibrium bulk modulus and equilibrium volume respectively.
$m_1$ and $n_1$ are parameters and are related by the following relations: $m_1 = 3(2n-1)$, $n_1 = 3(n-1)$, which are obtained by
imposing the volume analyticity condition. Since in this case the energy of the free atoms is zero, cohesive energy of the solid at $T=0K$ is
 the energy at which the $U(V)$ is minimum which happens to be at $V_0$. Thus the formula for cohesive energy $E_{coh}$ turns out to be

\begin{equation}
E_{coh} = \frac{B_0V_0}{(n-1)(2n-1)} \label{ecoh}
\end{equation}
\noindent
Also, it turns out that $n =\left. \frac{1}{3}\frac{dB}{dP}\right|_{V_0}$. The values of $B_0$, $V_0$ and $\left. \frac{dB}{dP}\right|_{V_0}$ are 
listed for various materials in the paper\cite{Jiu}. Cohesive energies calculated from the above formula are quite erroneous.
 Calculated values for some materials using Eq.(\ref{ecoh})  are compared with experimental values\cite{Kittel} in Table.\emph{1}.
 Also we compare the Energy per particle Vs Volume curve of Aluminum with
 the data obtained from ab-initio calculations\cite{Vasp} in Fig.(\ref{1}). It can be seen that there is a serious mismatch between the two. 
However from Fig.(\ref{1}), we can notice that the slopes of the mGLJ EOS and that of the ab-initio curve are similar which is the reason
 for pressure calculated from mGLJ EOS being accurate. 

\begin{table}
\caption{\label{11} Cohesive Energy } 
\begin{tabular}{c c c}  \hline
metal & \multicolumn{2}{c}{$E_{coh}(eV/atom)$}  \\ \cline{2-3}
&mgLJ EOS& $Experiment\cite{Kittel}$ \\ \hline\hline
 & &  \\
  Al& 9.02 &3.39 \\
  Cu&9.03 &3.49 \\
  Mo&59.1 &6.82 \\
   W& 86.4 &8.90 \\
   Zn&2.58 &1.34 \\ 
   Ag&5.05 &3.81 \\ 
   Pt&17.0 &5.84 \\ 
   Ti&102 &4.85 \\ 
   Na&5.68 &1.11 \\ 
   K&14.3 &0.93 \\ 
 & & \\ \hline
\end{tabular}
\end{table}

\section{mGLJ potential}
 The mGLJ potential is given by
\begin{equation}
u(r) = \frac{\epsilon}{3n}\left[ (3n-3)\left( \frac{r_0}{r}\right)^{6n-3}   - (6n-3)\left( \frac{r_0}{r}\right)^{3n-3}      \right] \label{glj}
\end{equation}
The parameters $\epsilon$, $n$ and $r_0$ are related to $B_0$, $V_0$ and $\left.\frac{dB}{dP}\right|_{V_0}$ denoted as $B_0\rq{}$ through the 
following relations.
\begin{eqnarray}
n = \left. \frac{1}{3}\frac{dB}{dP} \right|_{V_0} \\
r_0 = (\gamma V_0)^{1/3} \\
\epsilon = \frac{2B_0V_0}{(n-1)(2n-1)\delta}
\end{eqnarray}
where $\gamma$ is the structural constant which is $\sqrt 2$ for ${fcc}$ solids ans $2/\sqrt 3$ for ${bcc}$ solids and $\epsilon$ is
 the depth of the potential and $\delta$ is the number of first nearest neighbors. 
It can be seen that thermodynamic properties calculated using mGLJ potential with parameters of Sun diverge for materials with
$B_0\rq{}$ is less than $5$.
For example, consider the excess internal energy per particle ($U$) obtained through the
energy equation\cite{Hansen}.
\begin{equation}
U = 2\pi\rho \int_0^\infty u(r)g(r)r^2 dr \label{ee}
\end{equation}
where $\rho$ is the density of the system and $g(r)$ is the radial distribution function.
Since $g(r)$ becomes $1$ asymptotically, the integral requires that each term of $u(r)$ decays faster than $r^2$.
However, if $B_0\rq{}$
is less than $5$, the attractive component of $u(r)$ decays slower than $r^2$ allowing $U$ in Eq.(\ref{ee}) to 
diverge and for most of the materials $B_0\rq{}$ is less than $5$.
This renders the potential, as parameterized by Sun, to be inapplicable to calculate thermodynamic properties
 as they involve evaluation of integrals similar to Eq.(\ref{ee}). Also the potential cannot be used in
molecular simulations as the tail correction for internal energy is similar to Eq.(\ref{ee}) with lower limit being 
replaced by the cutoff radius of the potential.
\section{Conclusion}
 We noted that the mGLJ EOS predicts cohesive energies erroneously.
 Also we showed that the mGLJ potential cannot be used in liquid state theories and molecular simulations  for materials with $B_0'$ less than 
$5$ as the thermodynamic  quantities calculated using it diverge.
  This may be remedied by adjusting parameter $n$ so that $E_{coh}$ is properly reproduced.  
Also, including sufficient number of neighbors so that the total energy per particle converges would improve the results.
Lincoln et. al.\cite{Lin} obtained parameters of Morse potentials for various fcc and bcc materials by including up to $9^{th}$
neighbor shell. In a separate work, we have done the improvements mentioned above and obtained the parameters
by fitting the mGLJ EOS to ab-initio data. Same method is followed for EOS obtained from other pair potentials
 and the results are analyzed\cite{sai}.

\section{Acknowledgements}
I am thankful to Dr. Chandrani Bhattacharya, discussions with whom led to this paper. I thank Dr. N.K. Gupta for his encouragement.

\end{document}